# Thermal performance of GaInSb quantum well lasers for silicon photonics applications

C. R. Fitch,[1] G. W. Read,[1] I. P. Marko,[1] D. A. Duffy,[1] L. Cerutti,[2] J. Rodriguez,[2] E. Tournié,[2] and S. J. Sweeney[1, a)]

**AFFILIATIONS**

[1]Advanced Technology Institute and Department of Physics, University of Surrey, Guildford GU2 7HX, United Kingdom

[2]IES, Université de Montpellier, CNRS, F-34000 Montpellier, France

a)**Author to whom correspondence should be addressed:** s.sweeney@surrey.ac.uk

**ABSTRACT**

A key component for the realization of silicon-photonics are integrated lasers operating in the important communications band near 1.55 µm. One approach is through the use of GaSb-based alloys which may be grown directly on silicon. In this study, silicon-compatible strained $Ga_{0.8}In_{0.2}Sb/Al_{0.35}Ga_{0.65}As_{0.03}Sb_{0.97}$ composite quantum well (CQW) lasers grown on GaSb substrates emitting at 1.55 µm have been developed and investigated in terms of their thermal performance. Variable temperature and high-pressure techniques were used to investigate the influence of device design on performance. These measurements show that the temperature dependence of the devices is dominated by carrier leakage to the X minima of the $Al_{0.35}Ga_{0.65}As_{0.03}Sb_{0.97}$ barrier layers accounting for up to 43% of the threshold current at room temperature. Improvement in device performance may be possible through refinements in the CQW design, while carrier confinement may be improved by optimization of the barrier layer composition. This investigation provides valuable design insights for the monolithic integration of GaSb-based lasers on silicon.

The realization of optoelectronic integrated circuits (OEICs) requires an efficient, silicon-compatible electrically pumped laser operating above room temperature (RT). The indirect nature of silicon makes it unsuitable as an active region. While heterogeneous integration may currently be the most advanced approach in terms of device performance[1], the longer term goal is direct epitaxial growth of III-V lasers on silicon[2]. However, the lattice constant and thermal expansion coefficient mismatch and the polar/ non-polar interface between silicon and traditional III-V laser materials causes large defect densities, leading to inefficient and unreliable lasers. Progress has been made in overcoming these challenges through the use of GaAs based 1.3 µm quantum dot lasers on silicon[3,4] and Ga(NAsP)/GaP/Si quantum well (QW) lasers at 800-900 nm[5,6]. However, an alternative material system and approach is required for long-haul telecoms applications operating around 1.5 µm.

Sb-containing alloys are of interest for growth on silicon since dislocations tend to propagate parallel to the Si/III-V-Sb interface rather than into the active layers, allowing growth of high-quality active regions. GaInSb/GaSb composite quantum well (CQW) lasers have been grown by molecular beam epitaxy on 4°-off (001) silicon substrates emitting at 1.55 µm at RT in pulsed mode[7]; near 1.55 µm in continuous wave (c.w.) on GaSb near RT[8]; and more recently at 1.59 µm c.w. at RT on 6° miscut silicon[9]. However, further development is needed to address high threshold current densities ($J_{th}$) and temperature sensitivity[10,11].

To commercialize on-silicon devices, it is important to understand the efficiency limiting mechanisms of the equivalent active regions grown on GaSb. In this paper we report on the thermal properties of GaInSb CQW devices on GaSb substrates[8] and use a range of experimental techniques to identify the principal processes limiting device performance[12].

The test devices (A, B, C) illustrated in figure 1, consist of three compressively strained $Ga_{0.8}In_{0.2}Sb$ QWs. The $Al_{0.35}Ga_{0.65}As_{0.03}Sb_{0.97}$ barriers and $Al_{0.9}Ga_{0.1}As_{0.07}Sb_{0.93}$ cladding layers are lattice-matched to the GaSb substrate.

| Material | Thickness (nm) |
|---|---|
| p-GaSb(100) | 300 |
| Graded p-AlGaAsSb | 100 |
| p$Al_{0.9}Ga_{0.1}As_{0.07}Sb_{0.93}$ | 1000 |
| $Al_{0.35}Ga_{0.65}As_{0.03}Sb_{0.97}$ | 200 |
| $Ga_{0.8}In_{0.2}Sb$ | 3.6 (A), 4.8 (B), 6 (C) |
| $Al_{0.35}Ga_{0.65}As_{0.03}Sb_{0.97}$ | 20 |
| $Ga_{0.8}In_{0.2}Sb$ | 3.6 (A), 4.8 (B), 6 (C) |
| $Al_{0.35}Ga_{0.65}As_{0.03}Sb_{0.97}$ | 20 |
| $Ga_{0.8}In_{0.2}Sb$ | 3.6 (A), 4.8 (B), 6 (C) |
| $Al_{0.35}Ga_{0.65}As_{0.03}Sb_{0.97}$ | 200 |
| n$Al_{0.9}Ga_{0.1}As_{0.07}Sb_{0.93}$ | 1000 |
| Graded n-AlGaAsSb | 100 |
| n-GaSb(100) | substrate |

**FIG. 1.** Test device structures.

Devices B and C are CQWs formed by the insertion of one (B) and two (C) 0.45 nm $Al_{0.68}In_{0.32}Sb$ barriers within each $Ga_{0.8}In_{0.2}Sb$ QW. The insertion of AlInSb monolayers into the wider wells introduces additional confinement, counteracting the reduction in bandgap caused by the additional width. Further details of the fabrication of these devices is given in reference[8].

Pulsed electrical injection (500 ns, 10 kHz) was used to minimize current heating effects. Device characteristics were measured as a function of temperature from 40-300 K using a closed-cycle cryostat system. Figure 2(a) shows the facet output intensity ($L$) variation with current density ($J$) and temperature ($T$) for representative device B. Figure 2(b) shows the extracted $J_{th}$ variation with temperature for all three devices.

For these lattice-matched devices, defect-related recombination is assumed to be negligible and at low temperatures, where $J_{th}$ is low, other forms of non-radiative recombination are also assumed negligible. For an ideal QW laser at low temperatures the radiative component of threshold current density, $J_{rad} \propto T$ [13] and for our devices $J_{rad}$ can be seen to dominate threshold below ~150 K.

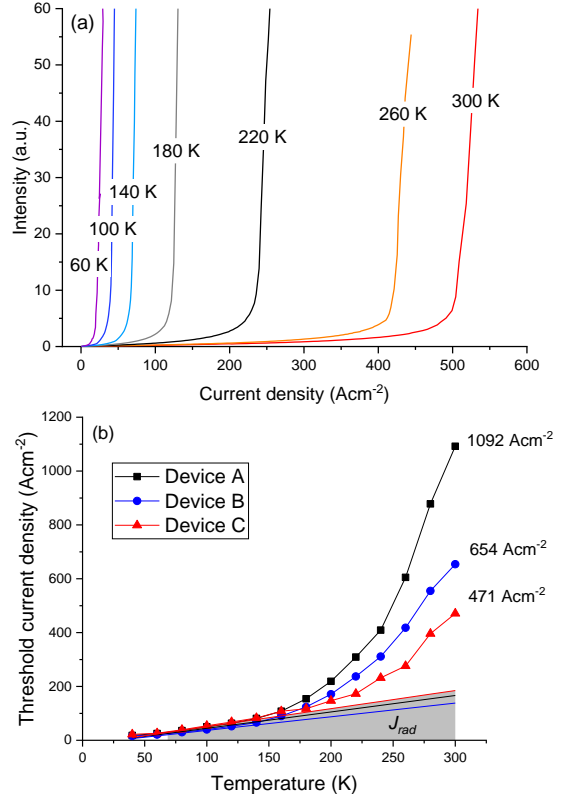

**FIG. 2.** (a) Temperature dependent LI (device B); (b) threshold current densities and radiative components (all devices).

The $J_{rad}$ components at RT were approximated by linear extrapolation of the low temperature $J_{th}$ to 300 K shown as the shaded area in figure 2(b) and in Table I.

**TABLE I.** Threshold current densities and radiative components at 300 K.

|  | Device A | Device B | Device C |
|---|---|---|---|
| $J_{th}$ (Acm$^{-2}$) | 1092 | 654 | 471 |
| $J_{rad}$ (Acm$^{-2}$) | 170±6 | 147±10 | 189±5 |
| $J_{rad}/J_{th}$ (%) | 15±1 | 23±2 | 40±1 |

For T > 150 K, $J_{th}$ increases super-linearly, suggesting the onset of non-radiative processes such as Auger recombination or carrier leakage consistent with evidence from other laser types at this wavelength[14].

In a simple model $J_{th}$ can be expressed as[15]:

$$J_{th} = eL_z(An + Bn^2 + Cn^3) + J_{leak} \quad (1)$$

Where $e$ is the electronic charge, $L_z$ is the active layer thickness, $n$ is the carrier density (assuming equal electron and hole carrier densities) and $A$, $B$ and $C$ are the recombination coefficients for defect, radiative and Auger recombination respectively. The $J_{leak}$ term accounts for carrier leakage from the QWs.

Auger recombination and carrier leakage are strongly temperature dependent which may explain the strong increase in $J_{th}$ with temperature[16,17]. The reduction in $J_{th}$ at RT from device structures A to B to C may be attributed to the increased gain volume, which reduces the threshold carrier density, $n_{th}$, by lowering the band filling and increasing the photon generation rate for a given injection current. Increased optical confinement would also contribute to the reduction in $n_{th}$ by increasing modal gain and the stimulated emission rate.

Approaching room temperature $J_{th}$ increases exponentially and, over a limited temperature range, this increase may be described by the characteristic temperature $T_0$[18]:

$$T_0 = \left(\frac{d\ln(J_{th})}{dT}\right)^{-1} = \left(\frac{1}{J_{th}}\frac{dJ_{th}}{dT}\right)^{-1} \quad (2)$$

A higher $T_0$ is desirable as it corresponds to greater thermal stability of $J_{th}$. Expressions for the characteristic temperature due to radiative, Auger recombination and leakage effects can be derived as[17,18]:

$$T_0(I_{rad}) = \frac{T}{1+2x} \quad (3)$$

$$T_0(I_{Aug}) = \frac{T}{3+\left(\frac{E_a}{kT}\right)+3x} \quad (4)$$

$$T_0(I_{leak}) = \frac{T}{\left(\frac{E_a}{kT}\right)} \quad (5)$$

Where $x$ is a "non-ideality" factor, e.g. due to optical losses, and $E_a$ is the respective activation energy for the Auger or leakage process.

The theoretical variation of $T_0(T)$ for $J_{th}$ can be written as a weighted average of the individual $T_0$ values:

$$\frac{1}{T_0} = \frac{1}{T_{0(rad)}}R + \frac{1}{T_{0(nonrad)}}(1-R) \quad (6)$$

where $R(T)$ is defined as $J_{rad}/J_{th}$ and $T_{0(nonrad)}$ corresponds to either the Auger or leakage process. $T_0(T)$ was measured using a five-point average and compared with a numerical model to investigate the non-radiative contribution to $J_{th}$.

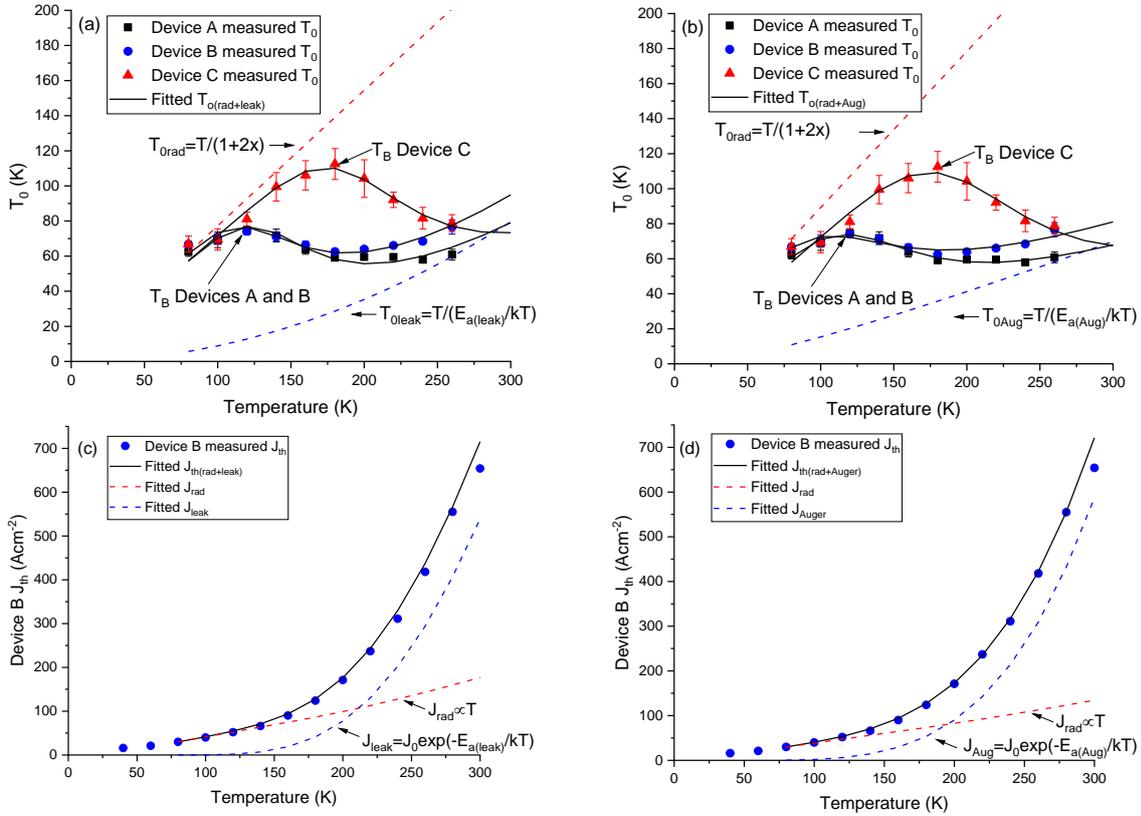

**FIG. 3.** Experimental and modelled characteristic temperature and threshold current density (a) $T_0$ (radiative and leakage) (b) $T_0$ (radiative and Auger) (c) $J_{th}$ (radiative and leakage) (d) $J_{th}$ (radiative and Auger).

Figure 3(a) shows the measured and modelled $T_0(T)$ for all three devices for radiative recombination and carrier leakage together with the radiative and leakage limits for device B as an example. Figure 3(b) shows that a similarly good fit may be achieved assuming radiative and Auger recombination. At low temperatures $T_0$ tends towards the radiative limits while above a break-point temperature, $T_B$[19], it tends towards the leakage or Auger limits. The difference in $T_0(T)$ between the three structures is consistent with the trends in $J_{th}$ and highlights the improvement in performance with increasing composite well thickness where a lower $n_{th}$ leads to a reduction in non-radiative recombination. Figures 3(c) and 3(d) show the example of the measured $J_{th}$ for device B and the result of fitting the radiative and leakage (a) or radiative and Auger (b) components from the $T_0$ model. A similar quality of fit is found for devices A and C. The $T_0$ data may therefore be explained by considering either leakage or Auger recombination although it is not possible to distinguish which is dominant from this analysis alone.

Using a similar approach we investigated the temperature sensitivity of the differential quantum efficiency (slope) above threshold $\eta_d$ where:

$$\eta_d \propto \frac{dL}{dI} \quad (7)$$

The characteristic temperature $T_1$ is defined as $\eta_d(T) = \eta_0 \exp(-T/T_1)$ where $\eta_0$ is a constant[20], so:

$$T_1 = -\left(\frac{d\ln(\eta_d)}{dT}\right)^{-1} = -\left(\frac{1}{\eta_d}\frac{d\eta_d}{dT}\right)^{-1} \quad (8)$$

Equation (8) was applied using a three-point average to plot the experimental values of $T_1$.

The differential quantum efficiency $\eta_d$ may be expressed as:

$$\eta_d = \eta_i \frac{\alpha_m}{\alpha_i + \alpha_m} \quad (9)$$

Here $\eta_i$ is the internal quantum efficiency, $\alpha_i$ is the internal loss and $\alpha_m$ the mirror loss expressed as:

$$\alpha_m = \frac{1}{L_{cav}} \ln \frac{1}{R} \quad (10)$$

where $L_{cav}$ is the cavity length and $R$ the as-cleaved facet reflectivity. Assuming the change in $\alpha_m$ with temperature is negligible we deduce that:

$$T_1 = \frac{1}{\frac{1}{(\alpha_i + \alpha_m)}\frac{d\alpha_i}{dT} - \frac{1}{\eta_i}\frac{d\eta_i}{dT}} \quad (11)$$

Terms on the left-hand side of the denominator relate to cavity losses and on the right-hand side to injection losses (e.g. carrier leakage). This decomposition allows investigation of the relative contributions to $T_1$. We used a least squares numerical fit model to vary $d\alpha_i/dT$ and $d\eta_i/dT$ to fit to the measured $\eta_d(T)$ using equation (11).

Figure 4(a) shows the relative change in slope efficiency $\eta_d$ for all three device structures. Figure 4(b) shows the measured and modelled average $T_1$ values. The limits associated with varying only $\alpha_i$ or $\eta_i$ with temperature are also shown.

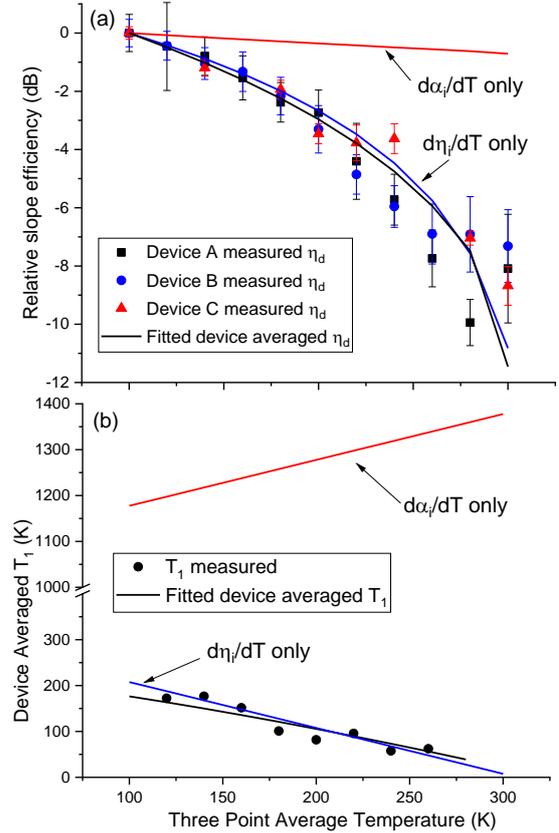

**FIG. 4.** Experimental and modelled slope efficiency (a) and characteristic temperature $T_1$ (b).

From this fit it is clear that $d\eta_i/dT$ dominates with $d\alpha_i/dT$ having negligible effect. This confirms that the temperature sensitivity of $T_1$ is due to injection rather than optical losses and supports the $T_0$ analysis which identified carrier leakage as a possible contributory factor. Auger recombination is not expected to influence the $T_1$ analysis since $n$ is (ideally) pinned above threshold. An important outcome of the $T_1$ analysis is that the efficiency data can only be explained by the presence of carrier leakage.

Both carrier leakage and Auger recombination are sensitive to band structure. Hydrostatic pressure can be used to reversibly change the band structure of a semiconductor, hence it was used to assess the dependence of the device performance on the band structure. Pressures up to 400 MPa were applied using a Unipress helium compressor system, details of which may be found in[16].

The measured variation of $J_{th}$ with pressure is shown in Figure 5.

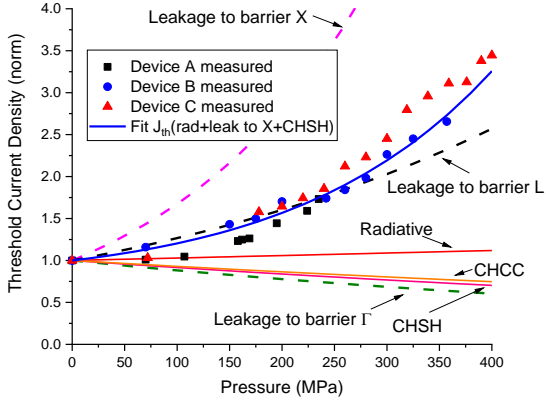

**FIG. 5.** Threshold current density variation with pressure - Experimental measurements (solid symbols) and theoretical dependence of the individual current paths and best fit with pressure (shown for device B).

A numerical model was created to describe how Auger recombination and carrier leakage vary with pressure to investigate their quality of fit to the experimental data.

In an ideal QW laser, where optical losses are negligible, $J_{rad}$ scales with bandgap according to $J_{rad} \propto E_g^2$.

The pressure dependence of the leakage and Auger components take the form:

$$J_{nonrad} = J_0 exp\left[-\left(\frac{dE_a}{dP}\right)\frac{P}{k_bT}\right] \quad (12)$$

Where $J_0 = 1$ for normalized data and $dE_a/dP$ is the net change in the respective activation energy with pressure.

For leakage, $dE_a/dP$ corresponds to the net change in band alignment between the quasi-Fermi level in the conduction band, taken to be equal to the QW electron ground state ($e1$), and the energy of the states into which the carriers escape. We assume that $e1$ changes as $dE_{lase}/dP$ and the valence band alignment is approximately pressure independent[21]. $dE_{lase}/dP$ were measured to be ~11 meVkbar$^{-1}$ for all three devices.

CHSH Auger recombination (where the energy of a Conduction band electron recombining with a hole in the Heavy hole band is given to a third hole in the Heavy hole band which is excited into the Spin split-off band) is sensitive to the difference between the band gap ($E_g$) and the spin-orbit (SO) split-off-energy ($\Delta_{SO}$). For our devices where $E_g > \Delta_{SO}$, the CHSH activation energy is given by:

$$E_a(CHSH) = \frac{m_{so}}{2m_h + m_c - m_{so}}(E_g - \Delta_{S0}) \quad (13)$$

where $m_c$ and $m_h$ are the electron and heavy hole band, in-plane, effective masses and $m_{so}$ is the SO split-off band effective mass.

The CHCC Auger activation energy is:

$$E_a(CHCC) = \frac{m_c}{m_c + m_h}E_g \quad (14)$$

CHCC refers to electron-hole recombination between the Conduction and Heavy hole bands accompanied by excitation of a Conduction band electron further into the Conduction band.

In (13) and (14) $E_g$ changes according to $dE_{lase}/dP$. From k.p theory, $m_c$ and $m_{so}$ increase approximately proportionally to band gap, and hence pressure, while $m_h$ is relatively independent of pressure[22]. QW effective masses were calculated from a linear interpolation of the binary components from Vurgaftman et al[23]. Other Auger processes have been ignored since they have previously been shown to be relatively weak[14,24]. Due to the closeness of the $e1$ and $\Delta_{SO}$ energies, CHSH is expected to dominate the Auger current at wavelengths < 2 μm[17].

An increasing $J_{th}$ with pressure is generally a strong indicator of carrier leakage to the indirect X or L valleys. In contrast, $J_{th}$ decreases with pressure for CHSH/CHCC Auger processes, or for leakage to the barrier Γ band edge (when the barrier Γ band edge pressure coefficient is smaller than the QW, as here).

The $e1$ to barrier Γ, X and L minima energy offsets were calculated at RT and ambient pressure for the three structures by combining the binary energy gaps, bowing parameters and valence band offsets from Vurgaftman et al[23], accounting for strain in the QW.

For the three structures, the range of values are: $e1$-$\Gamma_b$ = 218-223 meV; $e1$-$X_b$ = 275-280 meV; and $e1$-$L_b$ = 299-304 meV ($e1$-cladding separations were all >439 meV). While these energy offsets should be sufficient to confine the majority of carriers the Fermi-Dirac distribution of carriers extends into these energies where there is a high density of states, hence increasing the recombination rate. It was also found that the energy offsets and configuration of the barrier minima are very sensitive to the bowing parameters which vary considerably in the literature[23, 25].

As evident from Figure 5, the increase in $J_{th}$ with pressure is stronger than that expected for leakage to the $L_b$-valley states. However, it is consistent with leakage to the $X_b$-valley dominating the increase in $J_{th}$. A good fit to the experimental results was achieved for all devices using a combination of radiative recombination, leakage to the barrier $X_b$ minima (forming up to 43% of $J_{th}$), and a contribution of CHSH/CHCC Auger recombination or leakage to the barrier $\Gamma_b$ or $L_b$ minima.

The results of this study demonstrate the potential of GaSb-based QW lasers for silicon photonics applications in the telecoms wavelength range and opportunities for ongoing improvement.

We showed that the strong temperature sensitivity of $J_{th}$ could be explained by either leakage or Auger recombination. However, the temperature sensitivity of $\eta_d$ was found to be dominated by carrier injection, confirming that leakage must be present. Furthermore, from pressure-dependent measurements, carrier leakage, to the barrier $X_b$ valley states was found to be a key limiting factor in the performance of these devices. This could be reduced by increasing the activation energy of the leakage paths by a small increase in the lattice matched barrier Al and As fractions and without compromising optical confinement provided by the cladding. Optimization would also benefit from a detailed investigation of bowing parameters and band alignments in this system.

*Acknowledgments* This work has been partly supported by EPSRC (UK) under grant EP/N021037/1, a SEPnet PhD scholarship for D. A. Duffy, the French ANR (*Project OPTOSi*, No. ANR-12-BS03-002) and by the French "Investment for the Future" program (*EquipEx EXTRA*, No. ANR-11-EQPX-0016).

*Data Availability* The data associated with this work are available by request at https://doi.org/10.5281/zenodo.4353309.